\documentclass[onecolumn]{aa}

\usepackage{graphicx,natbib,hyperref,epsfig,lineno,longtable,txfonts}
\bibpunct{(}{)}{;}{a}{}{,}

\begin{document}

   \title{Magnetic helicity evolution during active region emergence and subsequent flare productivity}

   \author{Zheng Sun\inst{1}
          \and
          Ting Li\inst{2,3,4}
          \and
          Quan Wang\inst{2,3}
          \and
          Shangbin Yang\inst{2,3}
          \and
          Mei Zhang\inst{2,3}
          \and
          Yajie Chen\inst{5}
          }

   \institute{School of Earth and Space Sciences, Peking University, Beijing, 100871, People's Republic of China
         \and
         National Astronomical Observatories, Chinese Academy of Sciences, Beijing 100101, People's Republic of China\\
             \email{liting@nao.cas.cn}
         \and
         School of Astronomy and Space Science, University of Chinese Academy of Sciences, Beijing 100049, People's Republic of China
         \and
         State Key Laboratory of Solar Activity and Space Weather, Beijing 100190, China
         \and 
         Max Planck Institute for Solar System Research, Göttingen D-37077, Germany
             }

 
  \abstract
  {}
  {Solar active regions (ARs), which are formed by flux emergence, serve as the primary sources of solar eruptions. However, the specific physical mechanism that governs the emergence process and its relationship with flare productivity remains to be thoroughly understood.}
  {We examined 136 emerging ARs, focusing on the evolution of their magnetic helicity and magnetic energy during the emergence phase. Based on the relation between helicity accumulation and magnetic flux evolution, we categorized the samples and investigated their flare productivity.}
  {The emerging ARs we studied can be categorized into three types, Type-I, Type-II, and Type-III, and they account for 52.2\%, 25\%, and 22.8\% of the total number in our sample, respectively. Type-I ARs exhibit a synchronous increase in both the magnetic flux and magnetic helicity, while the magnetic helicity in Type-II ARs displays a lag in increasing behind the magnetic flux. Type-III ARs show obvious helicity injections of opposite signs. Significantly, 90\% of the flare-productive ARs (flare index $\geq$ 6) were identified as Type-I ARs, suggesting that this type of AR has a higher potential to become flare productive. In contrast, Type-II and Type-III ARs exhibited a low and moderate likelihood of becoming active, respectively. Our statistical analysis also revealed that Type-I ARs accumulate more magnetic helicity and energy, far beyond what is found in Type-II and Type-III ARs. Moreover, we observed that flare-productive ARs consistently accumulate a significant amount of helicity and energy during their emergence phase.}
  {These findings provide valuable insight into the flux emergence phenomena, offering promising possibilities for early-stage predictions of solar eruptions.}


   \maketitle
%

\section{Introduction}

Solar active regions (ARs) encompass all the observable occurrences that result from the expansion of the Sun's magnetic field, stretching from the photosphere to the corona, which become evident through emissions across a broad spectrum of wavelengths \citep{2015LRSP...12....1V,2019LRSP...16....3T}. Due to their strong magnetic field and complicated magnetic configuration, ARs are the principal source of a broad range of solar activities, ranging from small-scale brightenings and jets \citep{2010A&A...512L...2A,2019ScChE..62.1555C,2022ApJ...926L..39D,2023ApJ...953..148S} to large-scale flares and coronal mass ejections (CMEs) \citep{2015ApJ...804L...8L,2016A&A...591A.141J,2022ApJ...939L..18S}. Active regions are formed through the process of flux emergence from the convection zone in the Sun \citep{2010ApJ...720..233C,2001ApJ...561..406Z,2021LRSP...18....5F}. The initial stage of creating a simple bipolar AR involves the emergence of an $\Omega$-shaped flux tube. The interaction between the axial magnetic field of the tube and the photosphere leads to the emergence of two magnetic polarities within the AR. As the flux tube continues to emerge, the two main polarities gradually separate, giving rise to the appearance of small magnetic polarities between them \citep{2014LRSP...11....3C}. Despite numerous simulations having been conducted to study the process of AR emergence \citep{2014ApJ...789...35F,2017A&A...607A...1F}, the specific mechanisms of the emergence process and the correlation with subsequent flare and CME productivity remain subjects of ongoing discussions.

There are various parameters that can be utilized to characterize the properties of ARs \citep{2001ApJ...557L..71Z,2019ApJ...871...67C,2022ApJ...926L..14L}. \citet{2009ApJ...705..821W} classified these parameters into categories of extensive type and intensive type. Extensive-type parameters generally scale with the size of the AR, such as the magnetic flux $\Phi$ and magnetic energy $E$. On the other hand, intensive-type parameters are normalized quantities, for example, the reconnection flux divided by the total flux $\Phi_{r}/\Phi_{t}$ and free energy divided by the total energy $E_{f}/E_{t}$. The extensive-type parameters usually correlate with flares, while intensive-type parameters are associated with CMEs \citep{2015ApJ...804L..28S,2022ApJ...926...56K}. Several studies have demonstrated that the distribution of some parameters (e.g., magnetic shear, net current) near the polarity inversion line (PIL) can provide valuable insights into the productivity of solar eruptions \citep{2015ApJ...810...17D,2019SoPh..294..130K}. However, during the emergence of ARs, the PIL structure may not be unambiguously identified \citep{2012ApJ...753L..13S,2014SoPh..289.3351T}. Consequently, to investigate the properties of emerging ARs, it becomes necessary to consider parameters calculated over the whole region of ARs.
Theoretical studies simulating the magnetic flux tube emergence have suggested that a higher initial twist of the magnetic tube may result in a higher flux emergence rate $d(\Phi)/dt$ \citep{2001ApJ...549..608M,2006A&A...460..909M,2016SoPh..291..383F}. \citet{2017SoPh..292...48A} measured the flux emergence rate for 36 ARs and found that the relationship between $d(\Phi)/dt$ and the peak unsigned flux $\Phi_{p}$ could be described by a power-law index of 0.69 $\pm$ 0.10 for the entire sample of ARs. \citet{2021MNRAS.501.6076K} performed a statistical analysis and found a significant correlation between flare productivity and flux emergence rate, which offers valuable information for predicting flare productivity during the early phase of ARs.

In our study, we investigate the properties of emerging ARs by employing two key parameters: magnetic helicity and magnetic energy. Magnetic helicity serves as a measure of the net linking and the twisting of magnetic field lines within a given volume \citep{1999GMS...111....1B,2014SSRv..186..285P,2016SSRv..201..147V}. It is important to note that magnetic helicity is only gauge invariant under the condition that the normal component $B_{n}$ of the magnetic field vanishes on the bounding surface \citep{1958PNAS...44..489W,1999GMS...111....1B}. However, in the solar atmosphere, this condition is not satisfied due to the presence of a strong vertical magnetic component perpendicular to the solar surface. To address this issue, the concept of relative magnetic helicity is introduced \citep{1984JFM...147..133B,2013SoPh..283..369Y,2018A&A...613A..27Y}. Relative magnetic helicity is defined as the difference between the total magnetic helicity and the magnetic helicity of a reference field. This quantity maintains gauge invariance even when the boundary condition does not match, making it an effective characterization of helicity in the solar atmosphere. For the sake of simplicity, we refer to relative magnetic helicity as "magnetic helicity" throughout this paper. Regarding magnetic energy, it serves as a fundamental parameter that quantifies the amount of energy stored within magnetic fields \citep{2003SoPh..215..203D,2004Natur.430..326T}.

Various studies have demonstrated the relationship between magnetic helicity and solar eruptions \citep{2008ApJ...683.1160Z,2015RAA....15.1537J,2021ApJ...911...79P,2023arXiv230805366W}. \citet{2007ApJ...671..955L} conducted a statistical analysis of 48 X-flaring regions and 345 non-X-flaring regions, revealing that a peak helicity flux threshold of $6\times 10^{36}$ Mx$^2$s$^{-1}$ is required to produce an X-class flare. \citet{2010ApJ...718...43P} performed a statistical study of 378 ARs and found a strong correlation between helicity accumulation and flare productivity. They also observed a distinct difference in the helicity injection rate between flaring and non-flaring ARs. \citet{2020ApJ...897L..23K} and \citet{2022ApJ...925..129S} observed differences in the identified periodicities of magnetic helicity between flaring ARs associated with large M- and X-class flares compared to ARs associated with smaller B- and C-class flares.
When characterizing CMEs, the use of intensive-type parameters is generally considered more appropriate \citep{2015ApJ...804L..28S,2017ApJ...850...39T}. \citet{2017A&A...601A.125P} proposed the ratio of magnetic helicity of the current-carrying magnetic field to the total relative helicity $H_{j}/H_{r}$ as a useful indicator for CME eruptions. Numerous subsequent observational cases have supported this proposition \citep{2019A&A...628A..50M,2019ApJ...887...64T,2021A&A...653A..69G}. Among them, \citet{2021A&A...653A..69G} examined ten ARs and found that $H_{j}/H_{r}$ effectively distinguishes between CME-productive and CME-poor ARs. However, \citet{2023ApJ...945..102D} conducted a statistical analysis of 45 M- and X-class flares from 30 different ARs and found no systematic differences in the value of $H_{j}/H_{r}$ between confined and eruptive flares. The ratio of free magnetic energy to potential energy $E_{f}/E_{p}$ and normalized current-carrying helicity $H_{j}/\Phi^{2}$ instead exhibited the highest discriminative ability between confined and eruptive flares.

Numerous simulations and observations have investigated the transport of helicity from the solar interior to the atmosphere during flux emergence \citep{2003ApJ...593.1217P,2009A&A...502..333Y,2014ApJ...785...13L,2019JPlPh..85b7701P,2021A&A...652A..55W}. However, there is a lack of large-sample statistical analysis focusing on the helicity evolution during AR emergence. Moreover, there are only a few statistical investigations examining the relationship between emerging ARs and subsequent eruptions \citep{2021MNRAS.501.6076K,2022A&A...662A...6L}. Our work aims to examine the evolution of helicity in emerging ARs and subsequent AR flare productivity. We analyzed 136 samples that we divided into three categories based on the evolution of magnetic helicity and magnetic flux and found the differences in flare productivity among them. This paper is organized as follows. Section \ref{sec:dataset} and Section \ref{sec:methods} describe the dataset and analysis methods used in our study, respectively. In Section \ref{sec:Classification}, we divide the samples into three distinct categories according to the evolution of magnetic helicity and magnetic flux. Section \ref{sec:Results} presents our statistical results in detail. Finally, we discuss and summarize the major results in Section \ref{sec:discussion}.

\section{Dataset} \label{sec:dataset}
The data used in our study were obtained from the Helioseismic and Magnetic Imager (HMI; \citealt{2012SoPh..275..207S}) on board the Solar Dynamics Observatory (SDO), which provides magnetograms of the entire solar disk with a spatial resolution of 0.5$^{\prime\prime}$ pixel $^{-1}$. For our analysis, we mainly used the Spaceweather HMI Active Region Patch (SHARP) data \citep{2014SoPh..289.3549B}, which offers the advantage of automated identification and tracking of AR data. The SHARP data also comprises Lambert cylindrical equal-area (CEA) projections of the magnetic field vector, enabling the conversion of the components from [$B_{x}$, $B_{y}$, $B_{z}$] to the spherical heliographic components [$B_{\phi}$, $B_{\theta}$, $B_{r}$] with a cadence of 720s \citep{sun2013coordinate}, which can be used to calculate the magnetic helicity and magnetic energy. 

In total, we selected 136 emerging ARs according to the dataset provided in \citet{2021MNRAS.501.6076K}.
In the study conducted by \citet{2021MNRAS.501.6076K}, a total of 243 emerging ARs between 2010 May and 2017 December 31 were examined. The starting and ending times of the emergence phase were automatically determined using a two-segment piecewise continuous linear fitting method applied to the unsigned flux curves \citep{2019MNRAS.484.4393K}. The selection criteria for these samples are as follows: (i) clear identification of the starting and ending times of the emergence phase and (ii) ensuring that the ARs remain within the central meridian distance of less than 60$^\circ$ throughout the entire process. When we used SHARP to access these samples, in some cases the initial stages of emergence for some ARs were not detected by SHARP. We evaluated the ratio between the data-covering interval provided by SHARP and the overall emergence interval and considered that samples with a ratio above approximately 90\% can be regarded as possessing nearly complete temporal coverage. Given that there are seven samples with a ratio between 87\% and 90\%, we manually set the threshold as 87\% in order to include seven more samples and develop a more robust analysis. Consequently, based on the additional criterion we introduced, namely (iii) SHARP data has over 87\% time range of the emergence process. In this way, we selected 136 samples in total: 61 flaring ARs and 75 non-flaring ARs (shown in Table \ref{chartable}).

\section{Methods} \label{sec:methods}
We computed unsigned magnetic flux, magnetic helicity, magnetic energy and flare index for the selected 136 samples within their respective emergence time intervals:

For unsigned magnetic flux $\Phi$, we performed a summation of the absolute magnetic flux density in pixels on the radial magnetic field component $B_{r}$ provided by SHARP. Only pixels with magnetic flux density values exceeding 18 Mx cm$^{-2}$ were included in the calculation, corresponding to three times the noise level of the HMI magnetogram \citep{2012SoPh..279..295L}. We used the peak value (normally at the final time step) as the peak unsigned magnetic flux $\Phi_{p}$.

We employed a helicity flux method to calculate the magnetic helicity $H$ \citep{1999GMS...111....1B}. The helicity flux equation, Equation (1), is shown below:
\begin{equation}
  \left.\frac{d H}{d t}\right|_S=2 \int_S\left(\mathbf{A}_P \cdot \mathbf{B}_t\right) V_{\perp n} d S-2 \int_S\left(\mathbf{A}_P \cdot \mathbf{V}_{\perp t}\right) B_n d S.
\end{equation}
In the equation, $\mathbf{A_{p}}$ represents the magnetic vector potential, while $B_{n}$ and $\mathbf{B_{t}}$ denote the normal and tangential magnetic field, respectively. Similarly, $V_{n}$ and $\mathbf{V_{t}}$ respectively indicate the normal and tangential velocity. Typically, the integral surface $\mathbf{S}$ for the helicity flux calculation should be a closed hexahedron situated in the corona, with the photosphere serving as the bottom boundary. However, since the contribution of helicity is primarily provided by the photosphere, we considered the integral surface $\mathbf{S}$ to be the photosphere itself \citep{2014SSRv..186..285P}. The first term on the right-hand side of the equation is associated with $V_{n}$ and is referred to as the "emergence term." This term reflects the helicity contributed by the emergence of magnetic structures from the convection zone. The second term on the right-hand side is associated with $\mathbf{V_{t}}$ and is known as the "shear term." This term reflects the helicity contributed by the shearing motion of magnetic structures in the photosphere. 

To calculate the velocity vector on the photosphere, we used the differential affine velocity estimator for vector magnetograms (DAVE4VM; \citealt{2008ApJ...683.1134S}) method. We used an apodization window size of 19 pixels following the suggestion by \citet{2008ApJ...683.1134S}. In this method, the achieved velocity undergoes an additional adjustment to account for the exclusion of any irrelevant field-aligned plasma flow \citep{2012ApJ...761..105L,2022ApJ...929..122W}. This correction is achieved using the following equation:
\begin{equation}
  \mathbf{V}_{\perp}=\mathbf{V}-\frac{\mathbf{V} \cdot \mathbf{B}}{B^2} \mathbf{B}.
\end{equation}
Here, $\mathbf{V}_{\perp}$ represents the velocity perpendicular to the magnetic field lines, while $\mathbf{V}$ corresponds to the velocity obtained through the application of the DAVE4VM. The corrected velocity $\mathbf{V}_{\perp}$ serves as the crucial component for calculating the helicity flux. \citet{2012ApJ...761..105L} and \citet{2014ApJ...785...13L} also used DAVE4VM to calculate the helicity flux in emerging ARs. They used Monte Carlo methods to estimate the numerical errors when calculating the helicity flux, and the results are considered to be within the margin of errors. Thus, in this paper we regard the calculation as reliable.

Since we could calculate $dH/dt$ for each time step, we integrated $dH/dt$ over the entire duration of emergence. This integration allowed us to obtain the accumulated change in helicity, denoted as $\Delta H$: 
\begin{equation}
  \Delta H=\int_{t_{0}}^{t}{\frac{dH}{dt}}dt.
\end{equation}
To evaluate the eruptive potential of ARs based solely on their structural complexity rather than their area, it is valuable to examine the helicity normalized by the square of the magnetic flux \citep{2022arXiv220406010T}, denoted as $\Delta H$/$\Phi_{p}^2$.

We employed the energy flux method, equation (4), 
\begin{equation}
  \left.\frac{d E}{d t}\right|_S=\frac{1}{4 \pi} \int_S B_t^2 V_{\perp n} d S-\frac{1}{4 \pi} \int_S\left(\mathbf{B}_t \cdot \mathbf{V}_{\perp t}\right) B_n d S,
\end{equation}which is similar to the helicity flux method, in order to calculate magnetic energy $E$. . Once we obtained the energy flux dE/dt for each time step, we integrated these values to obtain the accumulated energy, denoted as $\Delta E$:
\begin{equation}
  \Delta E=\int_{t_{0}}^{t}{\frac{dE}{dt}}dt.
\end{equation}

Flare index (FI) was introduced by \citet{2005ApJ...629.1141A} to assess the flare productivity of an AR, which provides a quantitative measure of the AR's ability for flaring. The equation to calculate the FI is presented below:

\begin{equation}
  \mathrm{FI}=\left(100 S^{(\mathrm{X})}+10 S^{(\mathrm{M})}+1.0 S^{(\mathrm{C})}+0.1 S^{(\mathrm{B})}\right) / \tau.
  \end{equation}
  
Here, $S^{(j)}=\sum_{i=1}^{N_j} I_i^j$ represents the summation of the GOES peak intensities $I_i^j$ for a specific flare class (X, M, C, B), where $N_j$ denotes the number of flares within that class. In the above equation, $\tau$ signifies the total duration of the AR observation in days, indicating that this parameter is normalized by time. The FI data of the ARs are obtained from online dataset\footnote{\url{http://solar.dev.argh.team}}. It should be noted that the parameters $\Phi$, $H$, and $E$ are calculated during emergence, while FI includes the entire period of an AR until we could no longer observe it.

\section{Classification of active regions} \label{sec:Classification}
Once we had calculated the parameters outlined in Section 3, namely, the unsigned magnetic flux, helicity, and energy, we could construct the corresponding curves during the emergence phase. From the 136 samples, we found that almost all the unsigned magnetic flux $\Phi$ curves exhibit a monotonically increasing trend. This outcome is expected since the magnetic flux naturally increases as the AR emerges.
In contrast, the accumulated magnetic helicity $\Delta H$ curves manifested different characteristics. Based on the $\Delta H$ and $\Phi$ curves, we categorized the samples into three distinct groups. Figure 1 illustrates three representative examples for each category. It is worth mentioning that the intensity-normalized helicity represented by the orange curves refers to the normalization of helicity values into the range of (0, 1) rather than the previously mentioned normalized helicity $\Delta H$/$\Phi_{p}^2$.

The Type-I ARs (the first row in Figure 1) exhibited a simultaneous growth in the unsigned magnetic flux $\Phi$ and accumulated magnetic helicity $\Delta H$. The absolute value of the Pearson correlation coefficient between them exceeds 0.9. This high correlation indicates that the helicity steadily increases with unsigned flux, and it can be inferred that the helicity injection is mainly dominated by a single sign.
For Type-II ARs, the helicity curves exhibit a lag behind the $\Phi$ curve during the initial phase of emergence. As depicted in the middle row in Figure 1, the helicity curves demonstrate a slow initial growth followed by a rapid increase. As a result, the absolute value of the correlation coefficient between them is not more than 0.9. It is important to note that the second example of Type-II ARs displays a consistent helicity decrease. This is because this AR is dominated by a negative helicity. We also calculated the ratio between the lag period and the entire period for each Type-II AR, and the results showed that they exhibit a 50\% lag phase on average. In this way, the missing data (up to 13\% time period) of some samples would not affect the classification for Type-II ARs.
Type-III ARs display a significant injection of opposite helicity, resulting in a clear reversal of the helicity curve. Type-I and Type-II ARs may also exhibit slight opposite helicity injections in addition to their dominant helicity sign. However, they are relatively slight in contrast to the total helicity accumulation. To classify an AR as a Type-III, positive and negative signs of helicity must both be considered to constitute at least 30\% of the total unsigned helicity. 
Based on the criterion mentioned above, there are 71 Type-I ARs, 34 Type-II ARs, and 31 Type-III ARs, accounting for 52.2\%, 25\%, and 22.8\% of the total sample, respectively.

Figure 2 presents an example of $B_{r}$ magnetograms of AR 11620 corresponding to Type-I ARs, which provides a visual depiction of the entire process of AR emergence. During the emergence of the AR, a typical scenario involves the presence of a pair of positive and negative magnetic poles. As the AR emerges, the magnetic flux associated with both poles gradually increases, and the poles start to separate from each other. In the area where the poles meet, the PIL structures develop and are characterized by a significant magnetic field gradient and strong shearing motions. The clockwise rotation of the magnetic footpoints at the poles is also observed. This rotation contributes to the shearing term of the helicity injection within the AR. As the emergence process continues, the magnetic field becomes progressively more complicated. Apart from the two main magnetic poles, numerous smaller magnetic poles are also formed. The interaction of opposite polarities leads to ubiquitous magnetic cancellation. Figure 3 depicts the temporal evolution of the calculated parameters corresponding to Figure 2. In the top row, panel (a), of Fig. 3, we observed that the unsigned flux $\Phi$ and accumulated energy $\Delta E$ exhibited a monotonic increase over time. Panel (b) displays the time evolution of helicity flux $dH/dt$, while panel (c) presents the accumulated helicity $\Delta H$ obtained by integrating $dH/dt$. Notably, we can observe the strong correlation between the $\Delta H$ and $\Phi$ curves, a characteristic associated with Type-I ARs. Moreover, we observed that the shear term contributes to the majority of the helicity.

Figure 4 presents another example of AR 11768, classified as a Type-II AR, and illustrates its emergence from the quiet region. Through these magnetograms, we could observe a similar emergence process with Type-I ARs: a bipole magnetic configuration with a PIL structure in the middle. We could also observe the main pole rotation and the emerging smaller magnetic poles. As a result, it is hard to distinguish Type-I and Type-II ARs solely based on the magnetogram. Figure 5 displays the temporal evolution of the calculated parameters for this AR. We observed a monotonic increase in the $\Phi$, while the $\Delta H$ shows a lag behind $\Phi$ by about 30 hours. Moreover, helicity does not begin to increase rapidly until around 60 hours after the start of emergence. We found that the when the helicity is not accumulated prominently, the energy accumulates slowly, indicating a small amount of energy injection during the initial phase of emergence for Type-II ARs. When examining the magnetograms from the early phase of emergence (i.e., helicity accumulation is small), we could see a clear increase in magnetic flux (see Figure 4 and Figure 5(a)). Consequently, the minimal helicity injection can be attributed to the relatively diminished velocity field during the initial phase of emergence.

Figure 6 shows another example of AR 12521, and its temporal evolution of the calculated parameters is shown in Figure 7. This example is a Type-III AR characterized by a distinct reversal in the helicity curve. This implies that the helicity injection is primarily driven by positive helicity in the initial stage followed by a transition to negative helicity dominance. From Figure 6, we could not get conclusive evidence substantiating helicity reversal. The emergence process exhibits similarities with that of the other two types of ARs, rendering their differentiation based solely on magnetograms a challenge. However, the spatial area and the time interval during emergence are relatively smaller than the other two types. Importantly, discerning the source position of helicity reversal proved to be intricate due to the gauge-dependent nature of helicity flux density, which lacks a distinctly defined physical interpretation \citep{2005A&A...439.1191P,2023ApJ...942...27L}. We observed that the energy curve also displays alternating positive and negative phenomena, though the trend is different from the helicity. This phenomenon may be attributed to magnetic cancellations occurring between opposite polarities.
Notably, among all the cases, the shear terms (represented by the red curves) contribute significantly to the helicity, as opposed to the emerging terms (represented by the green curves). This indicates that the helicity injection is mainly caused by shearing motions rather than emerging twisted structures, which aligns with previous observations \citep{2014ApJ...785...13L,2015ApJ...806..245V,2022A&A...662A...6L}.

\section{Statistical results} \label{sec:Results}
Figure 8 displays the scatter plots on various parameters for the three types of ARs. 
Panel (a) shows the distribution of accumulated helicity $\Delta H$. It is evident that samples with markedly larger $\Delta H$ values (above $3\times10^{42}$ Mx$^2$) mainly belong to Type-I ARs. Furthermore, the $\Delta H$ values for Type-I ARs are significantly larger than those of the other two types. The mean $\Delta H$ for Type-I ARs is 1.81$\times 10^{42}$ Mx$^2$, approximately three times larger than that of Type-II ARs and about six times larger than that of Type-III ARs. 
Panel (b) presents the distribution of accumulated energy $\Delta E$. Notably, samples with significantly larger $\Delta E$ values (above $5\times10^{32}$ erg) are predominantly associated with Type-I ARs. Additionally, the $\Delta E$ values for Type-I ARs are also larger than those of the other two types. The mean $\Delta E$ for Type-I ARs is 1.51$\times 10^{32}$ erg, which is approximately twice as large as that of Type-II ARs and Type-III ARs. Panel (c) shows the distribution of normalized helicity. This quantity allowed us to assess the structural complexity of the ARs rather than focusing solely on their area \citep{2009AdSpR..43.1013D}. It is evident that Type-III ARs exhibit the lowest normalized helicity compared to the other two types of ARs.
In panel (d) the distribution of emergence time for the three types of ARs is presented. This parameter indicates the interval from the beginning of emergence to the end of the increase of the magnetic flux calculated by \citet{2019MNRAS.484.4393K}, as mentioned in Section 2. The analysis reveals that Type-II ARs exhibit a slightly longer mean emergence time compared to Type-I ARs. However, the ARs with a long emergence time, above 100 hours, are almost Type-I ARs. Notably, Type-III ARs have the shortest mean emergence time. This observation could be attributed to the fact that Type-III ARs typically have the smallest spatial range, as is evident from panels (a) and (b).
Panels (e) and (f) depict the distribution of helicity injection rate and the energy injection rate, respectively. Both panels highlight that Type-I ARs exhibit a higher injection rate, which is similar to the result in panels (a) and (b).
Based on the above statistical analyses, we could infer substantial differences among the three types of ARs. Moreover, $\Delta H$ and $\Delta E$ emerge as good indicators for Type-I ARs, while $\Delta H$/$\Phi_{p}^2$ can be used to distinguish Type-III ARs.

Figure 9 presents scatter plots related to the flare index and other parameters. Notably, panel (a) of this figure reveals that the vast majority (90\%) of ARs with a flare index greater than six belong to Type-I ARs. In contrast, only one sample (10\%) falls into the category of Type-III. In other words, Type-I ARs are the most likely candidates for becoming highly flare-productive ARs. Moreover, it is noteworthy that these highly flare-productive ARs consistently exhibit substantial helicity accumulation, with values exceeding 2.2$\times10^{42}\ {Mx}^2$. Furthermore, there are several Type-III ARs with a flare index above four, which indicates that Type-III ARs possess a moderate likelihood of becoming flare-productive ARs. In contrast, Type-II ARs demonstrate a notably limited propensity to develop into flare-productive regions.
In panel (b), a flare index trend similar to that in panel (a) can be seen. However, we observed that the normalized helicity $\Delta H$/$\Phi_{p}^2$ does not exhibit a strong correlation with the flare index. It appears that the flare index may not be closely associated with normalized helicity.
Panel(c) illustrates the $\Delta E$-$\Delta H$ diagram. A quasi-linear relationship with a Pearson coefficient of 0.8 can be observed between helicity and energy accumulation, which is consistent with previous observations \citep{2012ApJ...759L...4T,2022A&A...662A...6L}. Using the least-squares best logarithmic fits with a Kolmogorov-Smirnov significance level of about 0.8, the relationship between energy and helicity satisfies the following equation:  
\begin{equation}
  \Delta H=6.46\times 10^{11} \  \Delta E^{0.94}.
\end{equation}
Notably, samples with a high flare index (red and orange scatter points) tend to concentrate in the upper-right corner of the diagram, indicating that flare-productive ARs exhibit a larger helicity and energy accumulation.
However, non-flaring ARs (indicated by deep blue scatter points with FI=0) can also be found in the upper-right corner, suggesting that even with significant helicity and energy accumulation, certain ARs may remain inactive. Moderately active ARs (marked as green and light blue scatter points) can be observed in the lower-left corner, indicating that ARs with a relatively smaller helicity and energy injection can still exhibit moderate flare productivity. In contrast to the $\Delta E$-$\Delta H$ diagram with a random time duration in \citet{2012ApJ...759L...4T}, we strictly defined the starting and ending time of the emergence. In this way, the helicity and energy could reflect the exact value accumulated during the emergence of each AR. Moreover, we used data from SDO/HMI, which are widely used nowadays, to develop the helicity and energy calculation, thus providing a supplement to and validation of their work.

\section{Summary and discussion } \label{sec:discussion}
In this work, we selected 136 samples of emerging ARs and calculated their parameters of magnetic flux, helicity, and energy accumulation during emergence. Through analysis of the distribution of the parameters and their relationship with the flare index, we obtained the following important results:

(i) According to the time evolution of magnetic helicity and unsigned magnetic flux, we classified the emerging ARs into three categories. Type-I ARs exhibit a continuous increase in helicity during emergence  closely synchronized with the changes in magnetic flux, with a correlation coefficient above 0.9. In Type-II ARs, the helicity shows minimal growth during the initial phase of emergence followed by a rapid increase after a certain period, with a correlation coefficient below 0.9 compared to the magnetic flux. Type-III ARs display evident injection of opposite helicity, where the accumulated value of helicity with a specific sign constitutes at least 30\% of the total accumulated unsigned helicity.

(ii) The three types of ARs exhibit notable differences in terms of flare index, magnetic flux, helicity, normalized helicity, and energy. The most prominent distinction lies in the flare index. We observed that 90\% of highly flare-productive ARs ($FI \geq 6$) belong to Type-I ARs, suggesting that Type-I ARs are the most likely candidates for achieving a high level of flare productivity. In contrast, Type-II ARs show a very low likelihood of becoming flare productive, while Type-III ARs still maintain a moderate probability of becoming flare productive. Moreover, the $\Delta H$ and $\Delta E$ values for Type-I ARs are predominantly larger than those of the other two types, and significantly Type-III ARs have the lowest $\Delta H$/$\Phi_{p}^2$ value.

(iii) A highly flare-productive AR is certain to accumulate a significant amount of helicity and energy during its emergence phase. However, despite accumulating substantial energy during emergence, some ARs may remain as non-flaring. We also found that there is a quasi-linear relationship between helicity and energy accumulation, which is in accordance with previous studies \citep{2012ApJ...759L...4T,2022A&A...662A...6L}.

Our findings have two significant implications. Firstly, it provides valuable insights into the physical mechanisms of flux emergence from the perspectives of magnetic helicity and magnetic energy evolution. It also enhances our understanding of the transport of magnetic helicity from the solar interior to the solar atmosphere. Secondly, it helps establish a potential relationship between magnetic parameters during emergence and subsequent flare activities. Being able to identify whether an AR is flare productive or not when it emerges offers the possibility of early-stage predictions for solar eruptions.

The most distinct characteristic of Type-II ARs is their lag between the helicity and the unsigned flux.
They constitute 25\% of the total number of emerging ARs, indicating that the phenomenon of delayed helicity injection is widespread. 
This raises the question of why such a delay in helicity accumulation leads to a flare-poor nature. There are two possible explanations.
The first is that the coherency of the emerging flux tube is weak in Type-II ARs. At the initial stage of the emergence, because of the weak coherency of the flux tube, the magnetic helicity can be dispersed due to the convection flows at the photosphere, and thus there is only a small increase in the magnetic helicity. However, for Type-I ARs, which may have a strong coherency, the initial magnetic helicity is not dispersed due to the convection flows at the photosphere, and thus it accumulates from the beginning. In conclusion, the difference in the relations of magnetic helicity with magnetic flux for Type-I and Type-II ARs may reflect the coherency of the emerging magnetic flux tubes. The flux tubes with a strong coherency in Type-I ARs tend to produce more flares compared to those with a weak coherency in Type-II ARs. The second explanation is the lack or minimal injection of helicity during the initial phase of emergence. Previous studies as well as ours have shown that helicity injection is primarily driven by shear motions in the photosphere and sub-photosphere \citep{2014ApJ...785...13L,2022A&A...662A...6L}. The absence of helicity injection in the early stages of emergence suggests the absence of significant shear motions at the onset of emergence. \citet{2023NatSR..13.8994T} performed numerical simulations of the emergence of three types of magnetic flux tubes: non-twisted, weakly twisted, and strongly twisted. They placed the three magnetic flux tubes in the convection zone and observed their emerging process and helicity injection upward, respectively. Their results revealed that in the case of non-twisted flux tubes, the onset of helicity injection is delayed compared to the other two cases (see their Figure 2). The total helicity accumulation for the non-twisted case is also significantly less than the other two cases. Additionally, their model predicts that ARs formed by non-twisted flux tubes are hard to produce above M-Class flares. The scenario of the non-twisted case appears plausible when applied to Type-II ARs in our work. During the initial phase of emergence, the lack of prominent shear motions results in minimal helicity injection. A short time later, the footpoint motions induced by convective turbulence leads to the injection of helicity. These ARs formed by a non-twisted flux tube will have a relatively small helicity accumulation, and it will be difficult for them to be flare productive.

Type-III ARs demonstrate a distinctive characteristic of successive injections of opposite helicity signs. They constitute 22.8\% (31 out of 136) of the samples, but only one Type-III case can develop into an AR with a flare index greater than six. \citet{2006ApJ...644..575Z} proposed the concept of an upper bound on the total magnetic helicity that a force-free field can contain. Exceeding this upper bound would initiate a non-equilibrium situation, such as CMEs. In this scenario, the continuous injection of opposite helicity can effectively remove a portion of the accumulated helicity, thus inhibiting CMEs. To support this idea, \citet{2017A&A...597A.104V} and \citet{2021MNRAS.507.6037V,2022MNRAS.516..158V} conducted detailed case studies of several emerging ARs. They found that the helicity flux changed sign during the evolution of these cases and that the magnetic flux ropes and CMEs are absent. Though they are CME-poor, these cases can be associated with numerous C-class and sub-C-class flaring activities. They concluded that the injection of opposite helicity causes the relaxation of the sheared field through magnetic reconnection, resulting in small-scale flares, but it is difficult to form a large-scale flux rope and a large flare. Their investigations might explain why Type-III ARs have a moderate probability of becoming flare productive. Moreover, the injection of opposite magnetic helicity results in a smaller total helicity accumulation (see Fig. 8(d)), which is another contributing factor to the inhibition of large flares. There are indeed some studies showing that opposite helicity injection can cause a large flare \citep{park2012occurrence,2015RAA....15.1537J}; however, the detailed triggering mechanism between opposite helicity injection and flares is debated. These results may explain why four out of our 31 Type-III ARs can have a flare index exceeding four.

Through our study, we conducted a large-sample analysis of the relationship between parameters of emerging active regions and subsequent flare eruptions. However, several limitations must be acknowledged in our approach. Firstly, we did not consider the surrounding conditions of ARs during their emergence. In scenarios where an AR emerges in the presence of a pre-existing neighboring AR, the magnetic field structure can become more complicated, potentially leading to a higher probability of flare production \citep{2017ApJ...834...56T}. Thus, we focused solely on calculating the parameters of the AR within its emergence region. 
Secondly, we did not conduct a detailed analysis of how the energy and helicity accumulations trigger and contribute to flare eruptions. Our investigation only focused on a general examination of the relationship between the calculated parameters and the flare index. In future, we would like to investigate the long-term helicity evolution of ARs and their following solar eruptions in order to establish the relationship between AR helicity evolution and flare/CME productivity. Moreover, we will likely conduct some simulations focusing on the eruption potential of emerging ARs, which can provide possibilities for early-stage predictions of solar eruptions.

\graphicspath{{./}{figures/}}

\begin{figure}
   \centering
     \includegraphics[width=0.9\textwidth]{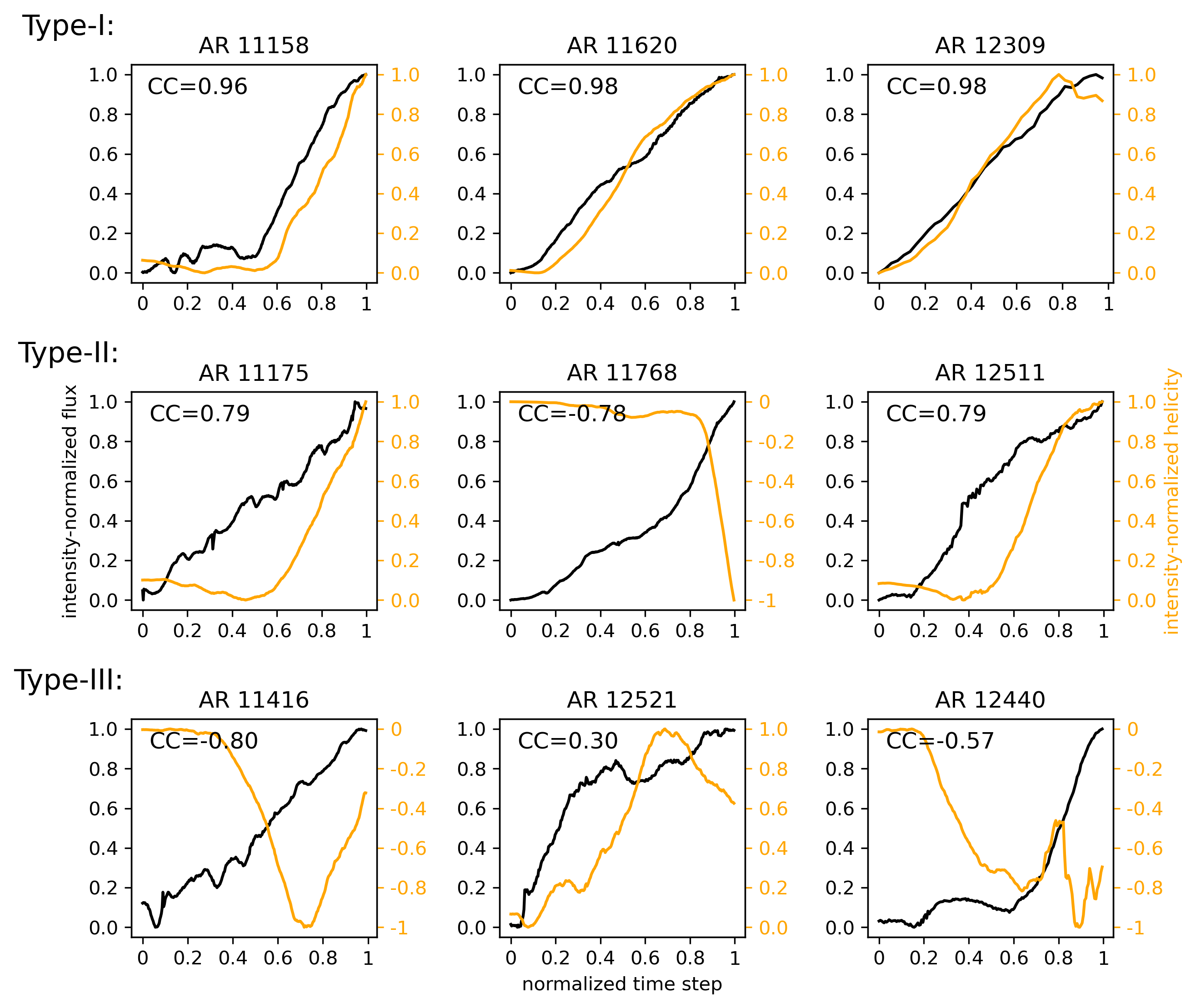}
     \caption{Unsigned magnetic flux $\Phi$ (blacks curve) and accumulated magnetic helicity $\Delta H$ (orange curves) evolutions for nine examples. The top, middle, and bottom rows represent the Type-I, Type-II, and Type-III ARs, respectively. The decreasing helicity indicates a negative helicity injection. The correlation coefficients between the two curves are marked in the upper-left corner of each panel.
   \label{fig:Fig.1}}
 \end{figure}
 
 \begin{figure}
   \centering
   \includegraphics[width=0.9\textwidth]{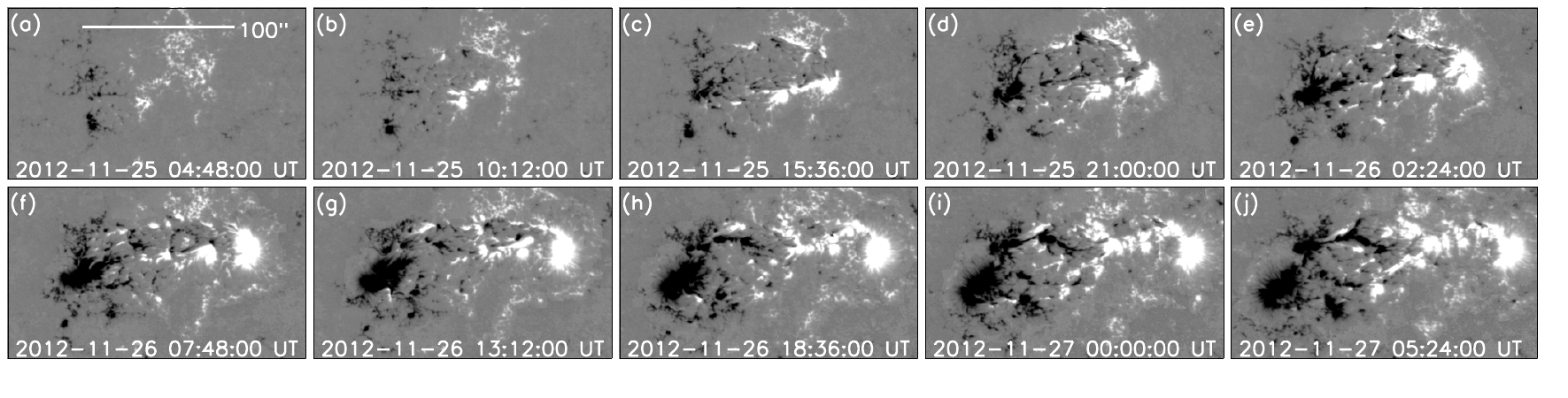}
   \caption{ Selected HMI images of the normal component, $B_n$
   , of the photospheric field of the Type-I AR 11620 showing the emergence process. The length of the horizontal white line corresponds to 100 arcsec.
   \label{fig:Fig.2}}
 \end{figure}
 
 \begin{figure}
   \centering
   \includegraphics[width=0.9\textwidth]{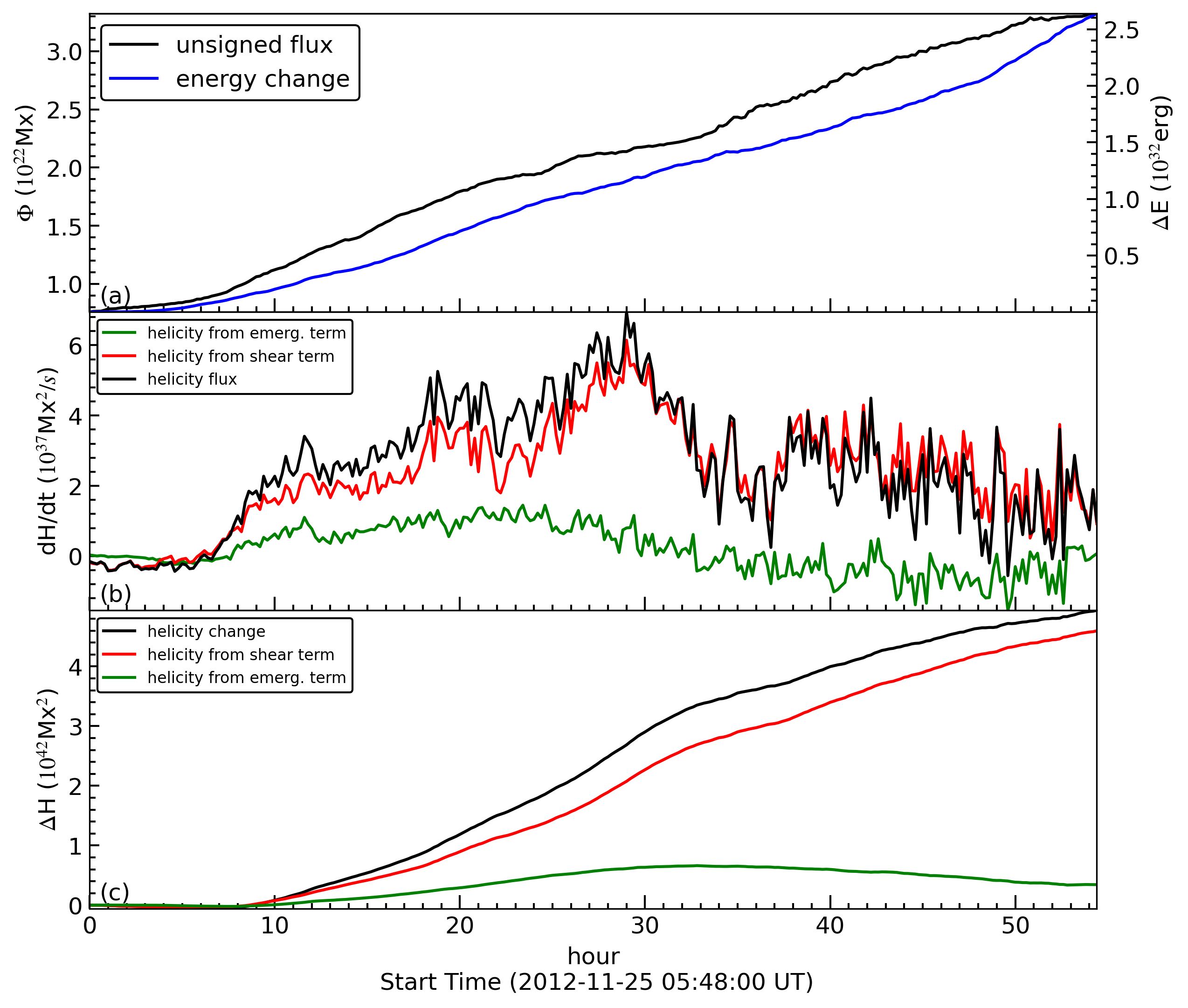}
   \caption{Time profiles of the magnetic flux, helicity, and energy for
  the Type-I AR 11620. Panel (a): Time evolution of the unsigned magnetic
   flux $\Phi$ (black) and accumulated energy $\Delta E$ (blue). Panel (b): Time profile of the helicity flux, $d(H)/dt$ (black). The red and green curves represent the shear and emergence terms, respectively. Panel (c):
   Time profiles of accumulated helicity $\Delta H$. The red and green curves indicate the helicity of the emergence term and shear term, respectively.
   \label{fig:Fig.3}}
 \end{figure}
 
 \begin{figure}
   \centering
   \includegraphics[width=0.9\textwidth]{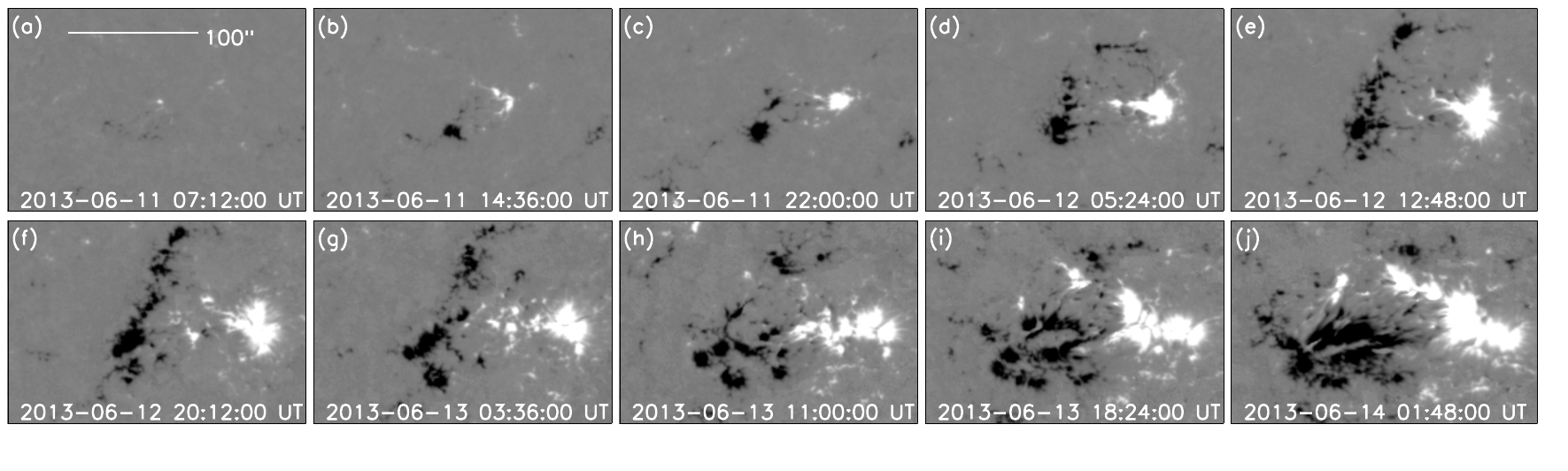}
   \caption{Same as Fig. 2 but for the Type-II AR 11768.
   \label{fig:Fig.4}}
 \end{figure}
 
 \begin{figure}
   \centering
   \includegraphics[width=0.9\textwidth]{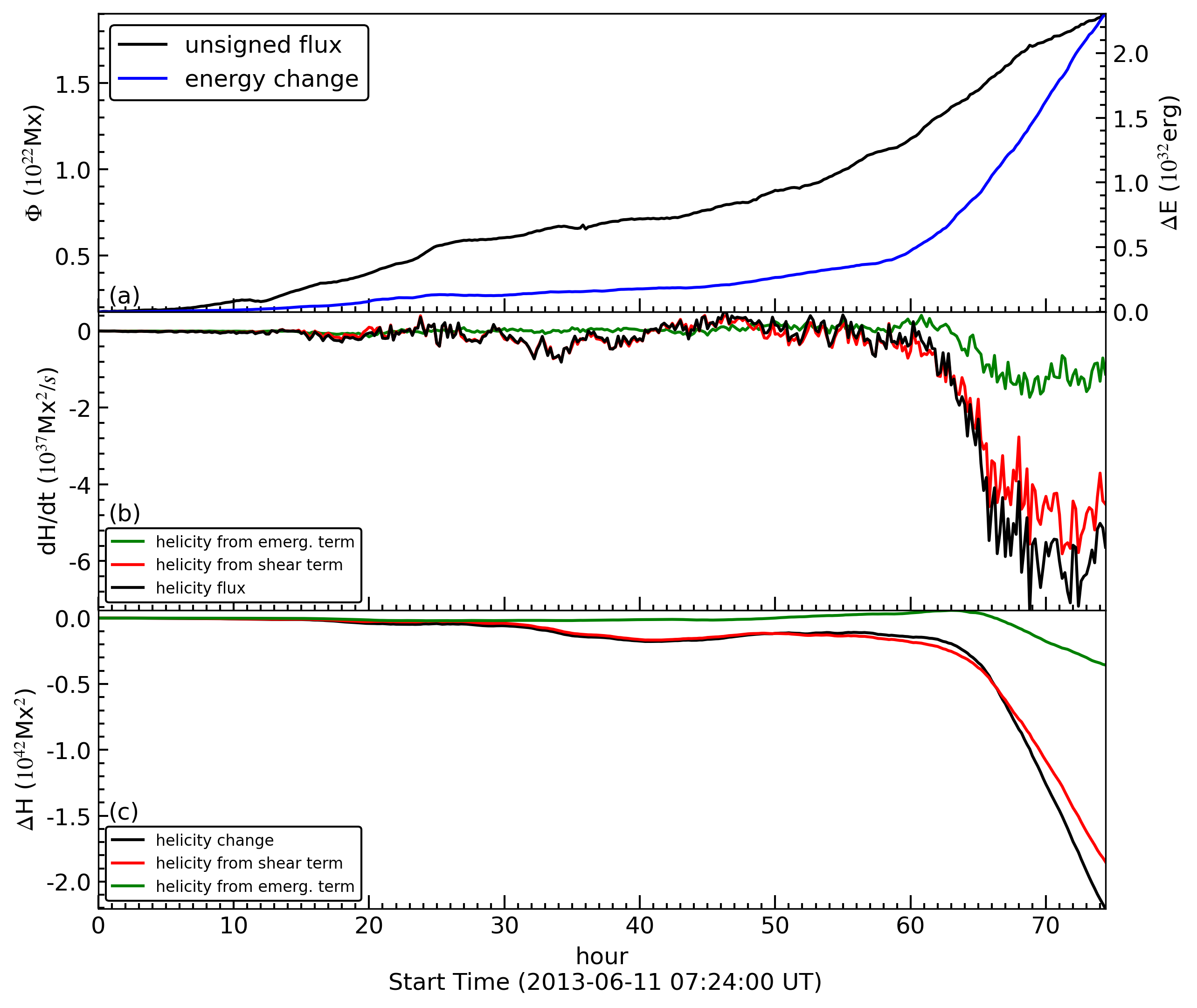}
   \caption{Same as Fig. 3 but for the Type-II AR 11768.
   \label{fig:Fig.5}}
 \end{figure}
 
 \begin{figure}
   \centering
   \includegraphics[width=0.9\textwidth]{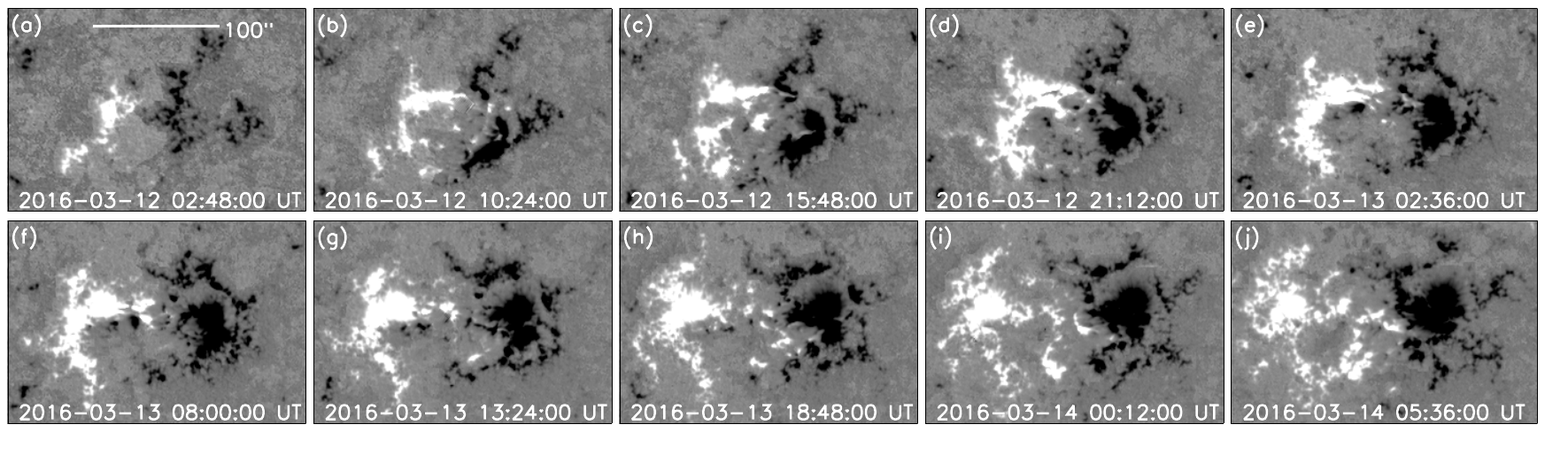}
   \caption{Same as Fig. 2 but for the Type-III AR 12521.
   \label{fig:Fig.6}}
 \end{figure}
 
 \begin{figure}
   \centering
   \includegraphics[width=0.9\textwidth]{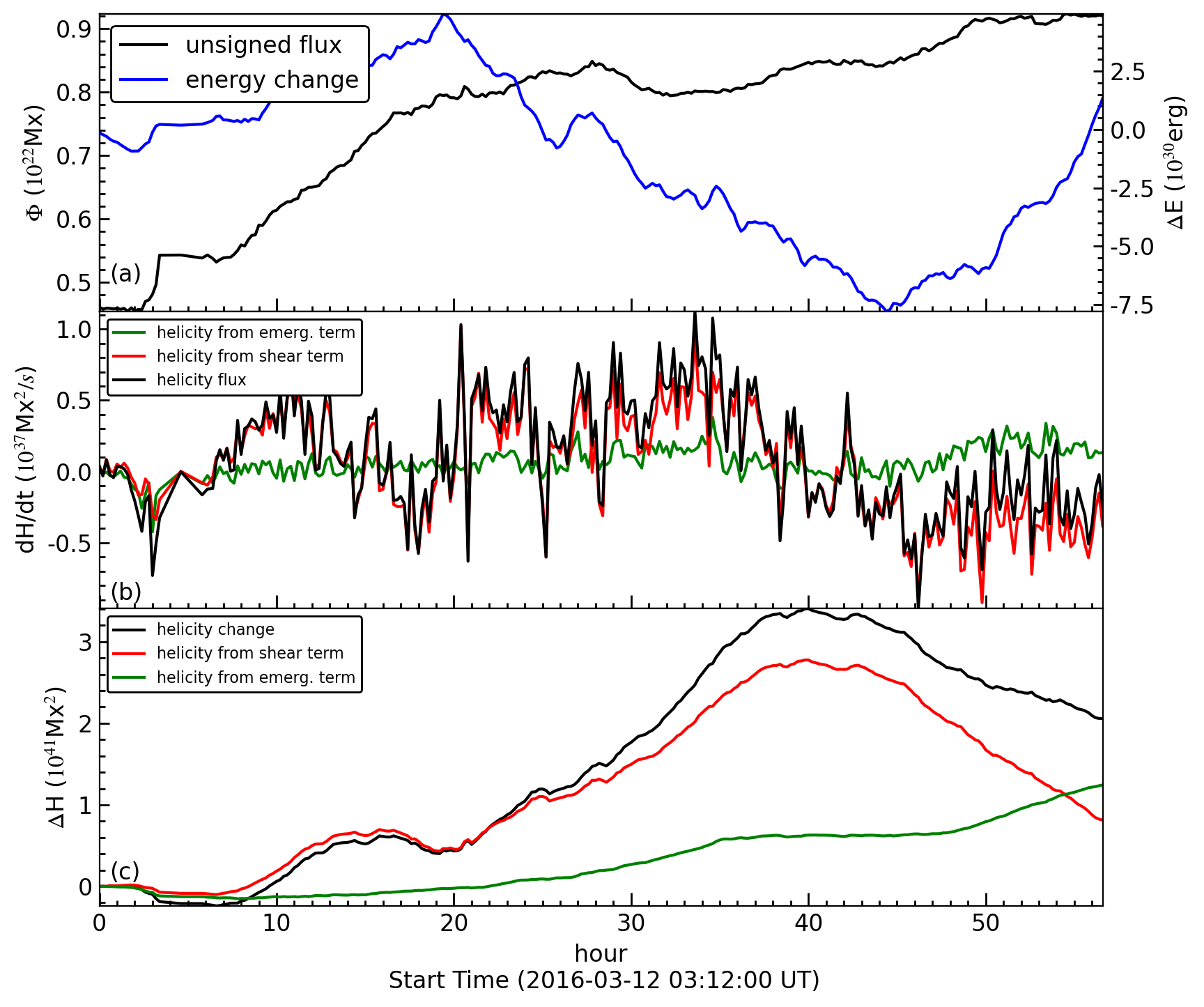}
   \caption{Same as Fig. 3 but for the Type-III AR 12521.
   \label{fig:Fig.7}}
 \end{figure}

 \begin{figure}
   \centering
   \includegraphics[width=1\textwidth]{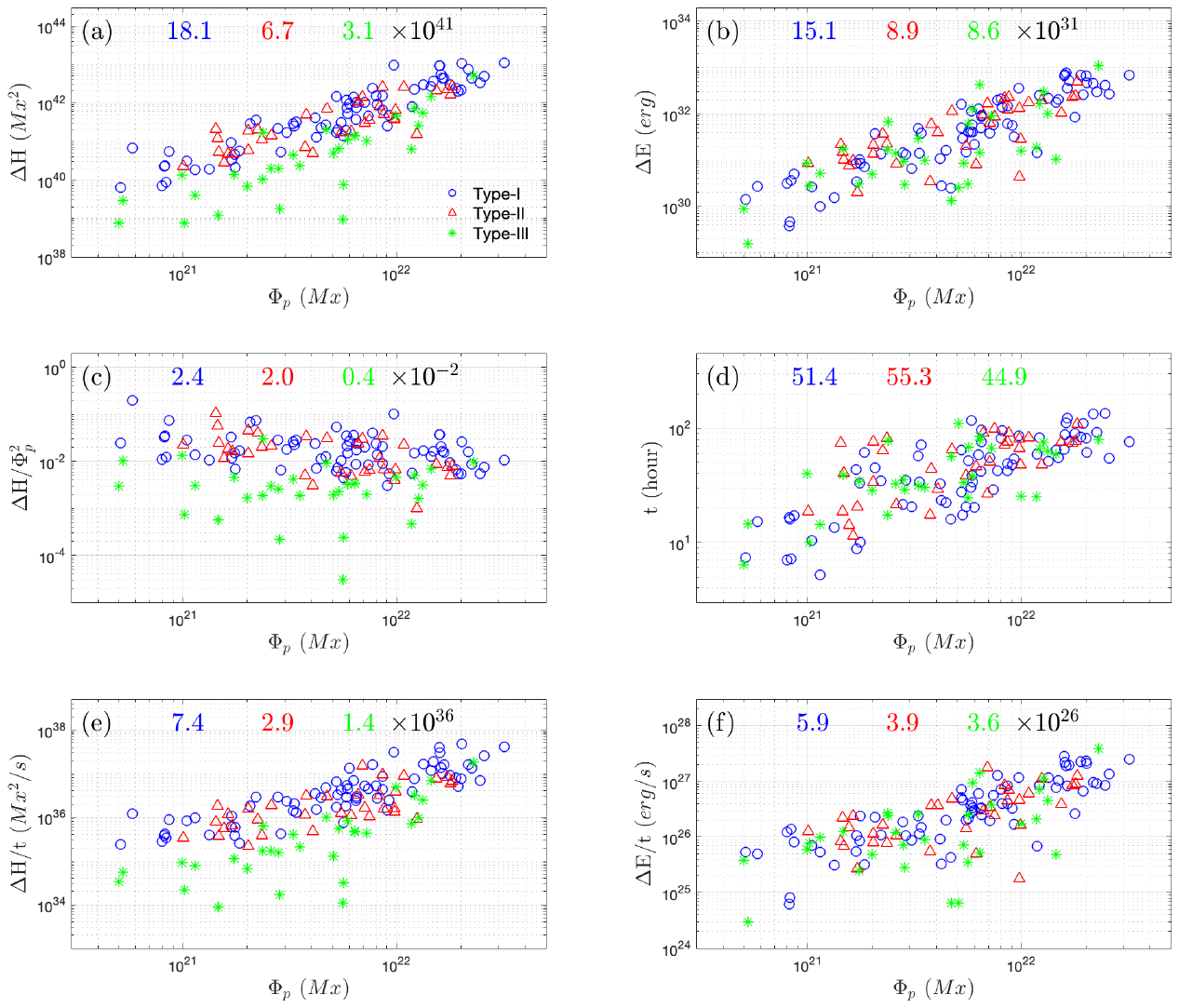}
   \caption{Scatter plots of various parameters for the three types of ARs. The horizontal axes represent the peak unsigned magnetic flux $\Phi_{p}$, while the vertical axes display the (a) accumulated helicity $\Delta H$, (b) accumulated energy $\Delta E$, (c) normalized helicity $\Delta H/\Phi_{p}^2$ , (d) emergence time $t$, (e) helicity injection rate $\Delta H/t$, and (f) energy injection rate $\Delta E/t$. The mean values of the parameters for the three types of ARs are listed at the top of each panel.
   \label{fig:Fig.8}}
 \end{figure}
 
 \begin{figure}
   \centering
   \includegraphics[width=1\textwidth]{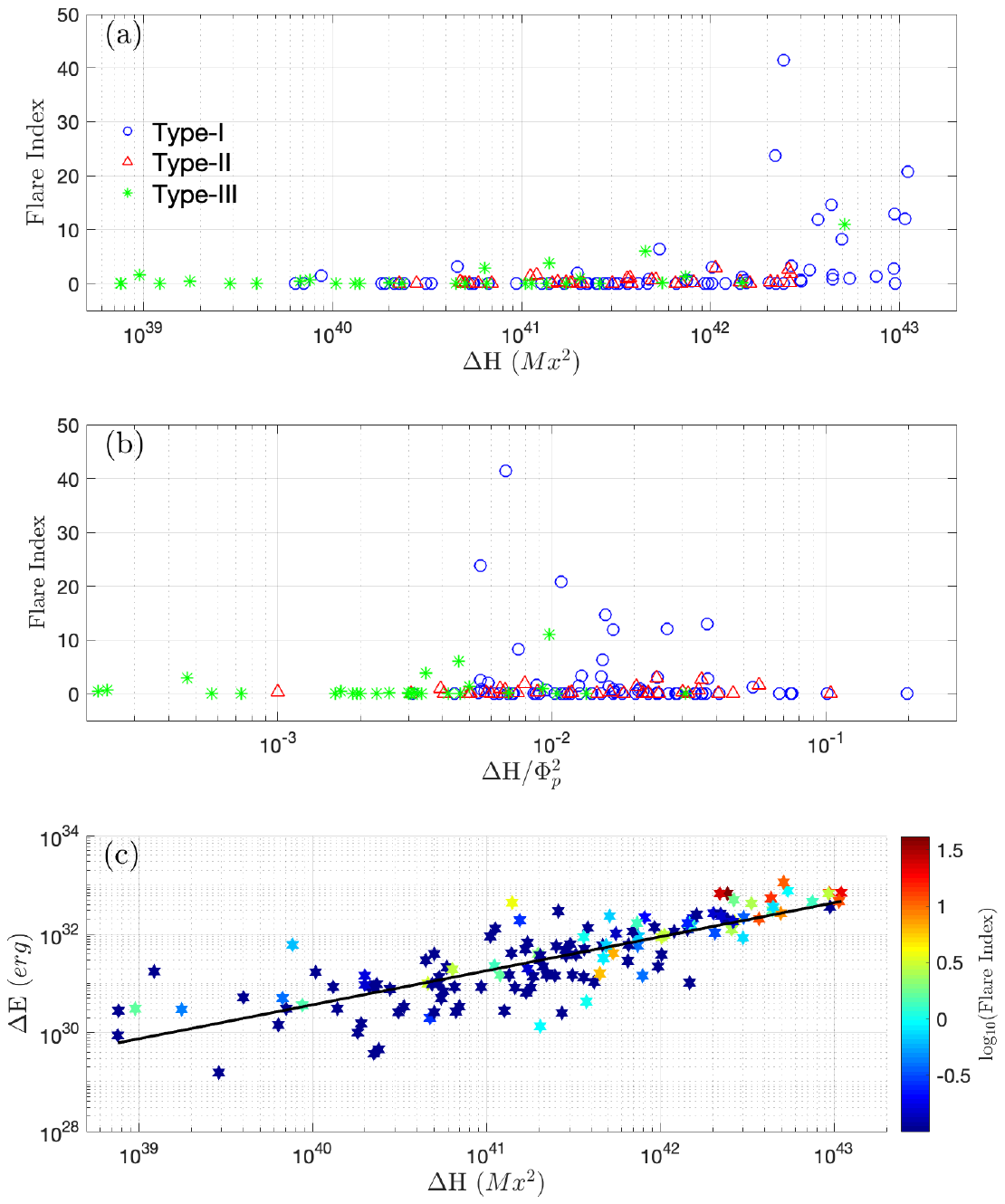}
   \caption{Scatter plots of the flare index versus the (a) accumulated helicity $\Delta H$, (b) normalized helicity $\Delta H/\Phi_{p}^2$, and (c) $\Delta H$-$\Delta E$ diagram. The color of the scatters in panel (c) indicate the value of the flare index. The black straight line represents the result of the least-squares best logarithmic fits between $\Delta E$ and $\Delta H$. 
   \label{fig:Fig.9}}
 \end{figure}

\begin{acknowledgements}
   We thank Dr. Jingxiu Wang and Dr. Shin Toriumi for fruitful suggestions. This work is supported by the Strategic Priority Research Program of the Chinese Academy of Sciences (Grant No. XDB0560000 and XDB41000000), the National Key R\&D Program of China (2019YFA0405000) and the National Natural Science Foundations of China (11973056, 12222306, and 12273060).
\end{acknowledgements}

\clearpage

\bibliographystyle{aa} 
\bibliography{ref}

\begin{appendix}
  \section{Calculated parameters of the dataset}
  \begin{longtable}{ccccccccc}
    \caption{Emerging AR samples and their corresponding parameters.\label{chartable}} \\
    \hline
    NOAA Number & Emergence Time & $\Delta H$ & $\Delta E$ & $\Phi_{p}$ & $\Delta H$/$\Phi_{p}^2$ & Correlation & Flare Index & Type \\
    & (hour) & $10^{41} (Mx^2)$ & $10^{31} (erg)$ & $10^{21} Mx$ & $10^{-2}$ & & & \\
    \hline
    \endfirsthead
    \multicolumn{9}{c}{{\tablename\ \thetable{} -- Continued from last page.}} \\
    \hline
    NOAA Number & Emergence Time & $\Delta H$ & $\Delta E$ & $\Phi_{p}$ & $\Delta H$/$\Phi_{p}^2$ & Correlation & Flare Index & Type \\
    & (hour) & $10^{41} (Mx^2)$ & $10^{31} (erg)$ & $10^{21} Mx$ & $10^{-2}$ & & & \\
    \hline
    \endhead
    \hline \multicolumn{9}{r}{{Continue...}} \\
    \endfoot
    \hline
    \endlastfoot
 
    11072 & 58.4 & 14.37 & 12.65 & 6.29 & 3.64 & 0.94 & 0.00 & I \\
    11076 & 80.6 & 3.82 & 13.30 & 9.72 & 0.40 & 0.77 & 0.00 & II \\
    11081 & 25.2 & 0.64 & 1.91 & 11.73 & 0.05 & 0.46 & 2.84 & III \\
    11096 & 42.0 & 3.65 & 4.96 & 6.46 & 0.87 & 0.95 & 0.00 & I \\
    11105 & 90.0 & 29.95 & 8.56 & 17.71 & 0.95 & 0.99 & 0.73 & I \\
    11130 & 74.6 & 2.60 & 29.79 & 12.66 & 0.16 & 0.83 & 0.00 & III \\
    11138 & 34.0 & 2.48 & 1.41 & 3.32 & 2.25 & 0.97 & 0.00 & I \\
    11141 & 38.6 & 7.39 & 5.83 & 5.86 & 2.15 & 0.96 & 0.38 & I \\
    11156 & 63.4 & 2.03 & 3.76 & 2.24 & 4.06 & 0.09 & 0.00 & II \\
    11158 & 83.2 & 24.26 & 67.76 & 18.89 & 0.68 & 0.96 & 41.46 & I \\
    11162 & 26.6 & 14.34 & 16.91 & 6.91 & 3.01 & 0.22 & 0.24 & II \\
    11175 & 77.1 & 16.12 & 24.82 & 17.87 & 0.50 & 0.79 & 0.00 & II \\
    11199 & 92.4 & 44.22 & 30.36 & 22.43 & 0.88 & 0.97 & 1.64 & I \\
    11214 & 95.6 & 9.23 & 13.80 & 8.12 & 1.40 & 0.98 & 0.00 & I \\
    11217 & 33.8 & 1.86 & 1.38 & 2.02 & 4.58 & 0.72 & 0.00 & II \\
    11223 & 31.4 & 0.45 & 2.93 & 3.27 & 0.42 & 0.54 & 0.00 & III \\
    11224 & 42.4 & 6.59 & 6.10 & 8.56 & 0.90 & 0.98 & 0.00 & I \\
    11241 & 57.0 & 2.85 & 4.83 & 3.27 & 2.66 & 0.96 & 0.00 & I \\
    11242 & 54.8 & 10.14 & 3.85 & 7.08 & 2.02 & 0.93 & 0.00 & I \\
    11273 & 34.2 & 0.14 & 0.31 & 1.73 & 0.46 & 0.20 & 0.00 & III \\
    11291 & 7.4 & 0.06 & 0.14 & 0.51 & 2.44 & 0.95 & 0.00 & I \\
    11297 & 25.4 & 4.50 & 1.59 & 9.91 & 0.46 & 0.50 & 5.98 & III \\
    11311 & 27.6 & 1.68 & 5.00 & 5.22 & 0.62 & 0.97 & 0.00 & I \\
    11318 & 32.8 & 0.20 & 1.42 & 2.58 & 0.30 & 0.56 & 0.17 & III \\
    11326 & 52.0 & 1.40 & 4.05 & 5.63 & 0.44 & 0.97 & 0.00 & I \\
    11344 & 90.4 & 10.59 & 9.66 & 6.62 & 2.41 & 0.73 & 2.89 & II \\
    11345 & 33.4 & 0.66 & 0.86 & 5.35 & 0.23 & 0.63 & 0.00 & III \\
    11365 & 50.6 & 3.03 & 6.31 & 7.11 & 0.60 & 0.37 & 0.00 & II \\
    11398 & 48.0 & 2.53 & 5.62 & 9.03 & 0.31 & 0.94 & 0.00 & I \\
    11404 & 14.4 & 0.04 & 0.51 & 1.14 & 0.31 & 0.44 & 0.00 & III \\
    11407 & 78.4 & 1.35 & 1.46 & 6.37 & 0.33 & 0.52 & 0.00 & III \\
    11416 & 59.4 & 14.70 & 1.03 & 14.53 & 0.70 & 0.99 & 0.10 & III \\
    11422 & 51.0 & 30.13 & 21.54 & 13.33 & 1.70 & 0.92 & 0.36 & I \\
    11430 & 76.0 & 25.85 & 12.39 & 8.57 & 3.52 & 0.63 & 2.67 & II \\
    11444 & 109.2 & 0.50 & 0.26 & 5.07 & 0.19 & 0.33 & 0.00 & III \\
    11446 & 40.6 & 0.53 & 1.00 & 1.47 & 2.45 & 0.48 & 0.00 & II \\
    11455 & 66.8 & 1.05 & 9.01 & 7.24 & 0.20 & 0.84 & 0.00 & III \\
    11456 & 14.6 & 0.03 & 0.02 & 0.52 & 1.04 & 0.02 & 0.00 & III \\
    11460 & 111.8 & 54.32 & 76.97 & 16.11 & 2.09 & 0.91 & 0.93 & I \\
    11464 & 6.4 & 0.01 & 0.09 & 0.50 & 0.30 & 0.54 & 0.00 & III \\
    11468 & 65.2 & 6.96 & 11.40 & 4.71 & 3.14 & 0.85 & 0.00 & II \\
    11473 & 17.4 & 2.03 & 2.92 & 5.30 & 0.72 & 0.92 & 0.00 & I \\
    11503 & 17.4 & 0.10 & 1.68 & 2.36 & 0.19 & 0.07 & 0.00 & III \\
    11523 & 116.4 & 11.99 & 11.43 & 5.83 & 3.52 & 0.93 & 0.00 & I \\
    11554 & 38.0 & 1.13 & 12.98 & 5.91 & 0.32 & 0.26 & 0.00 & III \\
    11560 & 85.6 & 94.95 & 36.09 & 9.69 & 10.10 & 0.95 & 0.00 & I \\
    11568 & 23.4 & 1.27 & 0.27 & 4.22 & 0.72 & 0.99 & 0.00 & I \\
    11620 & 54.4 & 49.47 & 26.38 & 25.63 & 0.75 & 0.98 & 8.22 & I \\
    11630 & 98.2 & 3.62 & 8.62 & 7.48 & 0.65 & 0.86 & 0.99 & II \\
    11632 & 110.4 & 20.29 & 26.26 & 19.38 & 0.54 & 0.97 & 0.15 & I \\
    11640 & 133.0 & 75.27 & 45.72 & 21.58 & 1.62 & 0.97 & 1.32 & I \\
    11670 & 82.0 & 15.47 & 21.25 & 8.00 & 2.42 & 0.96 & 0.15 & I \\
    11682 & 82.4 & 22.22 & 25.41 & 12.12 & 1.51 & 0.94 & 0.00 & I \\
    11697 & 30.4 & 0.23 & 0.98 & 3.52 & 0.19 & 0.03 & 0.00 & III \\
    11699 & 45.8 & 1.81 & 0.81 & 6.11 & 0.48 & 0.77 & 0.00 & II \\
    11706 & 39.8 & 0.13 & 0.85 & 0.99 & 1.34 & 0.87 & 0.00 & III \\
    11707 & 21.6 & 2.21 & 1.46 & 2.78 & 2.85 & 0.93 & 0.00 & I \\
    11709 & 61.2 & 0.56 & 0.71 & 1.83 & 1.67 & 0.92 & 0.00 & I \\
    11726 & 75.8 & 110.16 & 67.83 & 31.90 & 1.08 & 0.95 & 20.77 & I \\
    11762 & 61.6 & 106.61 & 46.75 & 20.12 & 2.63 & 0.95 & 12.00 & I \\
    11764 & 28.8 & 4.64 & 5.87 & 7.35 & 0.86 & 0.99 & 0.00 & I\\
    11765 & 71.6 & 8.13 & 22.00 & 8.31 & 1.18 & 0.85 & 0.21 & II\\
    11768 & 74.4 & 22.47 & 23.05 & 17.30 & 0.75 & 0.78 & 0.16 & II\\
    11776 & 65.6 & 7.35 & 16.81 & 12.13 & 0.50 & 0.71 & 1.29 & III\\
    11780 & 24.4 & 0.01 & 0.31 & 5.62 & 0.00 & 0.41 & 1.59 & III\\
    11781 & 62.8 & 5.54 & 10.14 & 13.28 & 0.31 & 0.89 & 0.14 & III \\
    11789 & 14.2 & 0.28 & 0.76 & 1.55 & 1.15 & 0.72 & 0.00 & II \\
    11831 & 47.8 & 6.52 & 2.83 & 9.89 & 0.67 & 0.89 & 0.00 & II \\
    11886 & 35.0 & 0.53 & 1.33 & 2.88 & 0.64 & 0.92 & 0.00 & I \\
    11889 & 75.6 & 7.46 & 9.13 & 6.51 & 1.76 & 0.92 & 0.72 & I \\
    11891 & 33.8 & 5.36 & 4.05 & 5.91 & 1.53 & 0.89 & 6.35 & I \\
    11902 & 43.4 & 0.93 & 0.86 & 1.68 & 3.31 & 0.96 & 0.00 & I \\
    11928 & 75.8 & 43.34 & 53.08 & 16.62 & 1.57 & 0.93 & 14.68 & I \\
    11933 & 20.2 & 9.69 & 2.21 & 6.03 & 2.67 & 0.96 & 0.00 & I \\
    11961 & 20.6 & 3.13 & 1.50 & 5.52 & 1.03 & 0.94 & 0.00 & I \\
    11978 & 13.6 & 0.19 & 0.15 & 1.33 & 1.08 & 0.96 & 0.00 & I \\
    11988 & 58.8 & 7.95 & 1.44 & 11.82 & 0.57 & 0.97 & 0.39 & I \\
    11992 & 34.8 & 3.61 & 1.35 & 2.19 & 7.51 & 0.98 & 0.00 & I \\
    11995 & 44.8 & 2.87 & 3.67 & 2.06 & 6.76 & 0.98 & 0.00 & I \\
    11999 & 18.6 & 0.23 & 0.84 & 1.00 & 2.25 & 0.78 & 0.00 & II \\
    12006 & 64.4 & 26.74 & 17.83 & 14.44 & 1.28 & 0.98 & 3.33 & I \\
    12018 & 77.6 & 1.73 & 6.79 & 2.37 & 3.07 & 0.12 & 0.00 & III \\
    12036 & 87.2 & 93.49 & 69.64 & 15.93 & 3.69 & 0.94 & 12.94 & I \\
    12037 & 65.0 & 10.81 & 16.78 & 5.81 & 3.32 & 0.94 & 2.80 & I \\
    12048 & 53.6 & 4.70 & 3.25 & 9.22 & 0.55 & 0.96 & 0.85 & I \\
    12051 & 55.2 & 37.10 & 20.62 & 14.90 & 1.67 & 0.96 & 11.90 & I \\
    12063 & 44.0 & 4.86 & 5.92 & 3.76 & 3.43 & 0.77 & 0.70 & II \\
    12085 & 79.4 & 51.27 & 111.36 & 22.89 & 0.98 & 0.78 & 11.02 & III \\
    12089 & 135.2 & 33.38 & 41.53 & 24.64 & 0.55 & 0.99 & 2.49 & I \\
    12091 & 21.6 & 1.45 & 0.82 & 2.60 & 2.15 & 0.86 & 0.00 & II \\
    12126 & 82.0 & 1.12 & 2.26 & 2.35 & 2.03 & 0.87 & 1.28 & II \\
    12129 & 11.4 & 0.48 & 0.97 & 1.63 & 1.83 & 0.89 & 0.00 & II \\
    12193 & 91.0 & 5.12 & 23.03 & 8.71 & 0.67 & 0.86 & 0.66 & II \\
    12198 & 10.4 & 0.31 & 0.26 & 1.04 & 2.88 & 0.97 & 0.00 & I \\
    12203 & 49.0 & 15.21 & 14.71 & 8.56 & 2.07 & 0.98 & 0.59 & I \\
    12219 & 109.0 & 26.57 & 49.36 & 18.26 & 0.80 & 0.84 & 1.80 & II \\
    12223 & 33.2 & 0.46 & 1.01 & 1.74 & 1.51 & 0.95 & 3.12 & I \\
    12226 & 66.3 & 3.75 & 0.43 & 9.77 & 0.39 & 0.80 & 1.00 & II \\
    12228 & 17.2 & 0.55 & 0.50 & 0.86 & 7.42 & 0.98 & 0.00 & I \\
    12229 & 15.2 & 0.67 & 0.27 & 0.58 & 19.80 & 0.97 & 0.00 & I \\
    12230 & 84.6 & 1.40 & 43.03 & 6.36 & 0.35 & 0.31 & 3.80 & III \\
    12231 & 18.6 & 1.20 & 1.49 & 1.45 & 5.70 & 0.73 & 1.54 & II \\
    12234 & 64.0 & 14.70 & 16.31 & 5.21 & 5.41 & 0.98 & 1.20 & I \\
    12257 & 81.2 & 21.95 & 67.38 & 20.02 & 0.55 & 0.94 & 23.79 & I \\
    12265 & 38.8 & 0.01 & 1.75 & 1.46 & 0.06 & 0.57 & 0.00 & III \\
    12266 & 42.6 & 24.28 & 19.91 & 7.73 & 4.07 & 0.90 & 0.00 & I \\
    12271 & 121.8 & 44.39 & 36.07 & 16.31 & 1.67 & 0.98 & 0.77 & I \\
    12275 & 47.6 & 1.55 & 19.14 & 12.45 & 0.10 & 0.69 & 0.26 & II \\
    12291 & 16.6 & 0.23 & 0.04 & 0.82 & 3.35 & 0.98 & 0.00 & I \\
    12294 & 30.4 & 1.98 & 4.00 & 5.80 & 0.59 & 0.91 & 1.93 & I \\
    12309 & 7.2 & 0.09 & 0.36 & 0.83 & 1.25 & 0.98 & 1.41 & I \\
    12315 & 29.2 & 0.50 & 3.90 & 4.04 & 0.31 & 0.12 & 0.00 & II \\
    12332 & 28.6 & 0.07 & 0.50 & 1.99 & 0.17 & 0.66 & 0.38 & III \\
    12353 & 68.0 & 0.08 & 6.16 & 5.66 & 0.02 & 0.45 & 0.63 & III \\
    12380 & 16.0 & 0.24 & 0.05 & 0.82 & 3.53 & 0.98 & 0.00 & I \\
    12388 & 28.8 & 0.02 & 0.30 & 2.83 & 0.02 & 0.67 & 0.45 & III \\
    12414 & 76.0 & 20.62 & 10.54 & 15.29 & 0.88 & 0.53 & 0.32 & II \\
    12424 & 8.8 & 0.33 & 0.34 & 1.69 & 1.17 & 1.00 & 0.00 & I \\
    12440 & 35.6 & 0.20 & 0.92 & 2.80 & 0.25 & 0.58 & 0.16 & III \\
    12453 & 75.8 & 0.59 & 2.15 & 2.02 & 1.46 & 0.22 & 0.00 & II \\
    12475 & 10.0 & 0.01 & 0.28 & 1.02 & 0.07 & 0.89 & 0.00 & III \\
    12500 & 20.6 & 0.47 & 0.20 & 1.70 & 1.63 & 0.87 & 0.26 & II \\
    12503 & 7.0 & 0.07 & 0.31 & 0.80 & 1.10 & 0.98 & 0.00 & I \\
    12509 & 10.0 & 0.21 & 0.85 & 1.76 & 0.69 & 0.90 & 0.00 & I \\
    12511 & 38.6 & 1.73 & 2.00 & 5.49 & 0.57 & 0.79 & 0.15 & II \\
    12514 & 22.2 & 2.23 & 1.61 & 4.43 & 1.14 & 0.97 & 0.00 & I \\
    12521 & 56.6 & 2.04 & 0.13 & 4.69 & 0.93 & 0.76 & 0.89 & III \\
    12528 & 71.4 & 3.26 & 3.82 & 3.39 & 2.83 & 0.95 & 0.00 & I \\
    12543 & 95.8 & 10.15 & 8.71 & 6.47 & 2.42 & 0.93 & 3.04 & I \\
    12548 & 83.0 & 26.25 & 17.92 & 10.82 & 2.24 & 0.89 & 0.16 & II \\
    12550 & 17.4 & 0.70 & 0.34 & 3.73 & 0.50 & 0.88 & 0.00 & II \\
    12557 & 74.4 & 2.11 & 2.20 & 1.42 & 10.41 & 0.75 & 0.00 & II \\
    12558 & 20.6 & 1.69 & 0.65 & 3.06 & 1.80 & 0.95 & 0.00 & I \\
    12607 & 16.0 & 2.73 & 0.25 & 4.65 & 1.26 & 0.94 & 0.00 & I \\
    12619 & 5.2 & 0.18 & 0.10 & 1.14 & 1.39 & 0.93 & 0.00 & I \\
    12632 & 32.8 & 4.16 & 1.07 & 4.15 & 2.41 & 0.93 & 0.00 & I \\  
  
  \end{longtable}
  \parbox{0.95\textwidth}{\small
  * The column "Emergence Time" shows the total time of emergence, which is calculated by the two-segment piecewise continuous linear fitting method applied to the unsigned flux curves \citep{2019MNRAS.484.4393K}. The columns of "$\Delta H$" and "$\Delta E$" indicate the total helicity and energy accumulation during emergence, respectively. The column of "$\Phi_{p}$" is the peak unsigned magnetic flux during emergence. The "Correlation" column indicates the Pearson correlation coefficient between the accumulated helicity $\Delta H$ curve and the unsigned flux $\Phi$ curve during emergence.}

\end{appendix}

\end{document}